\documentclass[12pt]{iopart}

\usepackage[dvips]{graphicx}

\begin{document}

\jl{2} 

\title{RF coupling of a Bose-Einstein condensate in a TOP trap}

\author{J L Martin, C R McKenzie, N R Thomas, D M Warrington and A C
Wilson}

\date{\today}

\address{Department of Physics, University of Otago, PO Box 56,
Dunedin, New Zealand}

\begin{abstract}
A radio frequency (RF) transition is used to convert a pure
$F=2,m_{F}=2$ $^{87}$Rb Bose-Einstein condensate confined in a TOP
trap to a mixture of $ F=2,m_{F}=2$ and $F=2,m_{F}=1$ states. We show
that the nature of this coupling process is strongly influenced by the
presence of the time varying field of the TOP trap, and complicated by
the presence of multiple Zeeman substates. In particular, the
effective Rabi frequency associated with the applied RF field is not
constant across the spatial extent of the cloud leading to a complex
geometry for atom-laser output coupling and `averaging out' of Rabi
oscillations. Further a time-varying detuning can give rise to complex
spatial structures.

\end{abstract}

 \submitted
\maketitle

\section{Introduction}
The development of atom-laser output coupling schemes and techniques
for coupling single state Bose-Einstein condensates (BEC) into
two-component condensates have been the subject of many recent
experimental studies.  The motivation behind these studies is the
development of practical output couplers and the investigation of
interactions within condensate mixtures.  For these experiments,
robust and controllable production of different output states is
essential. Two-component condensates have been produced using
sympathetic cooling \cite{JILA97}, two photon (RF and microwave)
transitions \cite{JILA98}, and RF coupling in optical dipole traps
\cite{MIT98}.  The later allows simultaneous trapping of all the
states within a hyperfine manifold. Several output coupling schemes
have been demonstrated experimentally: simple RF output coupling
\cite{MIT97,Hansch99,Otago99}, optically induced Bragg diffraction
\cite{Phillips99,MIT99} and Josephson tunneling \cite{Kasevich99}. Of
the six research groups who performed these studies, four used
time-averaged orbiting potential (TOP) traps \cite{JILA95} but none
considered effects introduced by the time-varying field.

This paper extends the study of RF output coupling to include dynamics
associated with the time-varying field of a TOP trap.  RF coupling in
a static trap was recently characterised by \cite{Minardi00}, and a
high degree of control demonstrated.  In a TOP trap, the population
dynamics are strongly influenced by the time-varying nature of the
trap. We find that the presence of the rotating magnetic field can
significantly alter the apparent spectral width of the applied RF
field, which in turn changes the spatial profile of the coupling
region.  We also find that the large quadrupole gradient of the TOP
trap gives rise to a large change in the effective Rabi frequency
across the cloud, leading to complex spatial population distributions.

\begin{figure}[tbph]
\begin{center}
\includegraphics[width=\textwidth]{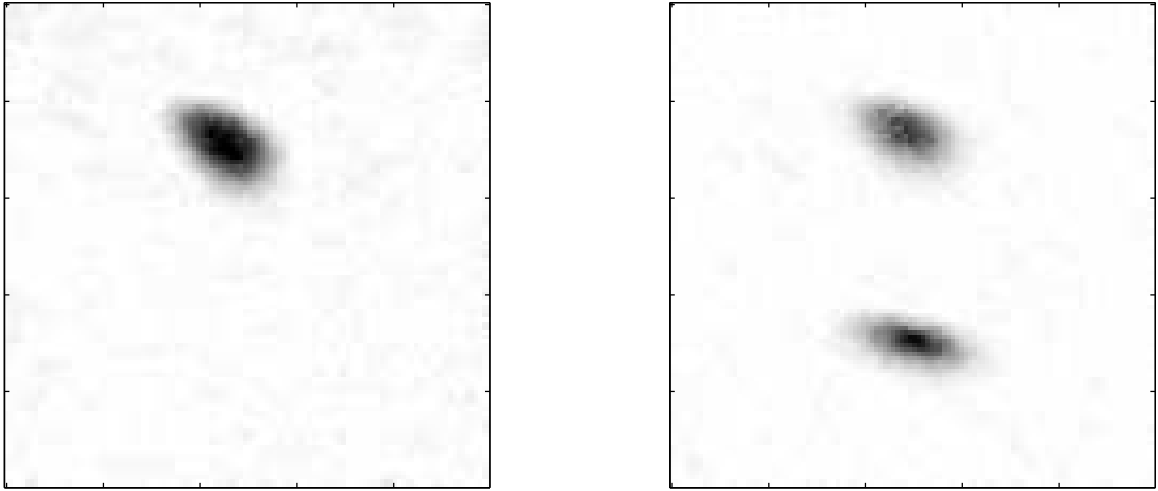} 
\end{center}
\caption{(a) An image of a Bose condensate without any coupling field
applied. (b) A similar condensate after application of an RF field
tuned to the centre of the trap with magnitude 72~mG applied for
32~$\mu $s.  The individual images are 250~$\mu $m along each axis and
are taken after a 5~ms time-of-flight through the inhomogeneous
magnetic field.}
\label{figure:twocomponent}
\end{figure}

Our experimental procedure has been described in \cite{Otago99}.  We
start with a $^{87}$Rb Bose-Einstein condensate in the $\left|
F=2,m_{F}=2\right\rangle $ state, confined in a TOP trap with radial
quadrupole field gradient $ B_{\mathrm{q}}^{^\prime}=200$~G~cm$^{-1}$
and bias field $B_{\rm TOP}=4$~G.  We then apply an RF coupling field
with polarization perpendicular to the plane of rotation of the TOP
field.  The effect of this field is to induce transitions between
adjacent $m_{F}$ states in the $F=2$ hyperfine manifold, and hence the
initially pure $\left| 2,2\right\rangle $ condensate state becomes a
mixture of the five possible $m_{F}$ states. By changing the
frequency, duration and magnitude of the coupling field different
admixtures of the manifold states can be generated. Note that states
$\left| 2,-1\right\rangle $ and $\left| 2,-2\right\rangle $ are high
field seeking and therefore actively repelled from the trap, and state
$\left| 2,0\right\rangle $ being essentially magnetically insensitive
is untrapped; hence these states will be present in the trapping
region for only a few milliseconds after the coupling pulse. Once the
coupling field has been applied we separate the trapped components by
removing the trapping potential and 1~ms later applying a magnetic
field gradient which introduces a Stern-Gerlach separation. The
different magnetic sub-states ($\left| 2,2\right\rangle $ and $\left|
2,1\right\rangle $) become easily resolvable after about
5~ms. Although the probing pulse is tuned on resonance with the
$\left| 2,2\right\rangle $ state, the population of the $\left|
2,1\right\rangle $ state is also accurately determined since the
50~$\mu $s probing pulse (in a 4~G bias field) leads to rapid optical
pumping into the $\left| 2,2\right\rangle $ state. This gives us
effective resonant probing of both condensate $m_{F}$ states. Typical
results of this process are shown in Fig.~\ref{figure:twocomponent}.

\section{Coupling in the TOP trap}
To analyze this coupling process, it is helpful conceptually to first
consider the case of a static magnetic trap, such as a Ioffe-Pritchard
(I-P) configuration. When an RF field is applied to couple atoms into
other states, the static magnetic field gradient gives rise to a
static but spatially varying detuning $\Delta $. To calculate the
linewidth of a pulse of duration $\tau _{\rm pulse}$ we must consider
the relative effects of the pulse spectral width and power broadening
(note that the natural linewidth of an RF transition is insignificant
in comparison). The pulse spectral width is given by
\[
\Gamma _{\rm spectral} \approx \frac{2\pi }{\tau _{\rm pulse}},
\]
and the power broadening (for $\left| 2,2\right\rangle \leftrightarrow
\left| 2,1\right\rangle $) by \cite{Loudon}
\[
\Gamma _{\rm power}=\frac{g_{F}\mu _{\mathrm{B}}B_{\rm
rf}}{\sqrt{2}\hbar },
\]
where $B_{\rm rf}$ is the magnetic field amplitude of the RF coupling
field. For the $\left| 2,1\right\rangle \leftrightarrow \left|
2,0\right\rangle $ transition the linewidth is a factor of
$\sqrt{2/3}$ smaller.

These two mechanisms combine to give an intrinsic linewidth associated
with the RF pulse. The detail of how these combine depends on the
exact form of each lineshape. The lineshape associated with power
broadening is Lorentzian. However, the spectral lineshape is
determined by the shape of the RF pulse, which for a square pulse
gives rise to a sinc spectral distribution. Numerical simulations
suggest that addition in quadrature is a good approximation, so that
\[
\Gamma _{\rm pulse} \approx \sqrt{ \Gamma _{\rm spectral}+\Gamma _{\rm
power} }.
\]

We now include the effects of a time varying magnetic field. In a
typical TOP trap significant angular change in the orientation of the
bias field will occur during even a short RF pulse (e.g. with a bias
field rotating at $\omega _{\rm TOP}$ = 7~kHz, a 10~$\mu $s pulse
corresponds to a 25$^{\circ}$ change in direction). The effect of this
rotation is to introduce time dependence to the already spatially
varying detuning. This time dependence affects the coupling process in
two distinct ways.  Firstly, as the quadrupole field rotates about the
centre of the cloud the detuning at any fixed point varies
sinusoidally with time. Secondly, this sinusoidal variation in the
detuning introduces a spatially dependent transit time broadening of
the transition linewidth.  Effectively, the available interaction time
is reduced from $\tau _{\rm pulse }$ to $\tau _{\rm transit}$ by the
time varying detuning. This transit time broadening, $\Gamma _{\rm
transit}$, depends on the distance $r$ from the axis of rotation of
the TOP field (in our case the z-axis) and $\Gamma _{ \rm pulse}$
\begin{equation}
\Gamma_{\rm transit} = 2\pi r\omega _{\rm
TOP}\frac{\omega_{\mathrm{q}}^{^\prime }} {\Gamma _{\rm pulse}},
\label{equation:gammatransit}
\end{equation}
where $\omega _{\mathrm{q}}^{^\prime}$ is the radial gradient of the
RF transition frequency (arising from the Zeeman shift associated with
the quadrupole field gradient $B_{\mathrm{q}}^{^\prime}$) given by
\begin{equation}
\omega _{\mathrm{q}}^{^\prime } =
g_{F}B_{\mathrm{q}}^{^\prime}\frac{\mu _{\mathrm{B}}}{\hbar }.
\label{equation:omegaqdash}
\end{equation}

By careful choice of experimental parameters, one might choose to
avoid the complications introduced by the rotating field. This could
be achieved, for example, by broadening the RF transition so that the
coupling rate is essentially the same for all atoms. In our
experiment, the transition frequency gradient is 14~kHz~$\mu $m$^{-1}$
and our condensate diameter is $\approx$ 30~$\mu $m, giving a change
in the transition frequency across the spatial extent of the cloud of
420~kHz.  Therefore, in order for the coupling to be insensitive to
the field rotation, the coupling field would need to have a linewidth
significantly larger than 420~kHz. This could be achieved in two ways,
with a very short pulse (duration $\ll 2 \mu $s) or a very intense RF
field (amplitude $\gg 0.85$ G).

\begin{figure}[tbph]
\begin{center}
\includegraphics[width=0.6\textwidth]{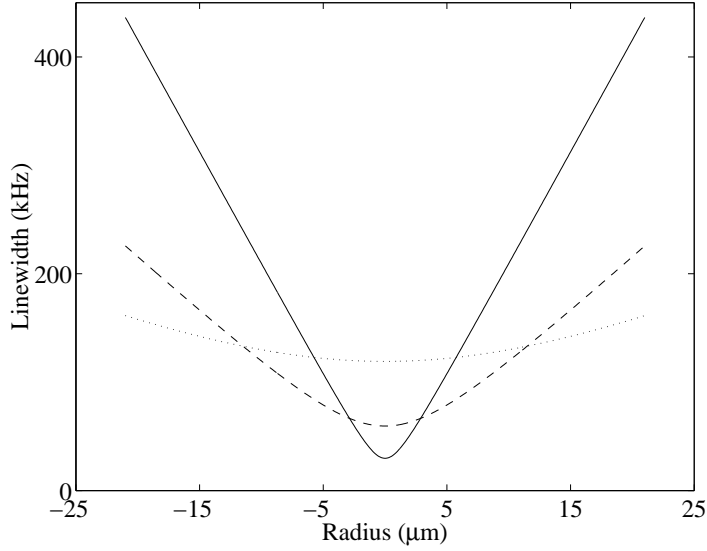} 
\end{center}
\caption{The total linewidth $\Gamma $ as a function of position
within the condensate for weak RF field and three different pulse
lengths: ---~32~$\mu $s, $-\!-\!-$~16~$\mu $s, and $\cdots$~8~$\mu
$s.}
\label{figure:linewidth}
\end{figure}

Using Eq. (\ref{equation:gammatransit}) we can now determine the total
RF transition linewidth
\[
\Gamma \approx \sqrt{ \Gamma _{\rm pulse}+\Gamma _{\rm transit} }.
\]
The spatial dependence of the linewidth is shown in
Fig. \ref{figure:linewidth} for three different values of $\Gamma
_{\rm pulse}$.

There are two regimes that simplify the expression for $\Gamma $. The
first of these is where $\Gamma _{\rm transit}\ll \Gamma _{\rm
pulse}$, in which case transit broadening can be ignored, and the
linewidth of the coupling process can be treated as a constant
(e.g. Figure \ref{figure:linewidth}, $\tau _{\rm pulse}=8$~$\mu $
s). The second case is when $\Gamma _{\rm transit}\gg \Gamma _{\rm
pulse} $ (e.g. in the case of quasi-continuous output coupling). From
Eq. (\ref {equation:gammatransit}) we see that the linewidth in this
case is approximately linear with radius. However, in our experiments
neither of these simple approximations apply and a more complete
treatment is required.

\section{Modeling the coupling}
We are able to broadly predict the population dynamics of the various
$m_{F}$ states as a function of time with the following simple Rabi
cycling model that neglects any kinetic or mean-field effects within
the condensate.  In an $F=2$ manifold there are four Rabi frequencies
$\Omega _{1\ldots 4}$ associated with the RF field, where $\Omega
_{1}$ couples states $\left| 2,2\right\rangle \leftrightarrow \left|
2,1\right\rangle $, $\Omega _{2}$ states $\left| 2,1\right\rangle
\leftrightarrow \left| 2,0\right\rangle $, etc. From the
Clebsch-Gordan coefficients for the transitions, $\Omega _{1}=\Omega
_{4}$, $\Omega _{2}=\Omega _{3}$ and $\sqrt{3}\Omega
_{1}=\sqrt{2}\Omega _{2}$ allowing us to define a single $\Omega (t)$
for all transitions
\[
\Omega (t)=\frac{g_{F}\mu _{\mathrm{B}}B_{\rm rf}(t)}{\sqrt{2}\hbar }.
\]

Each RF transition will be detuned an amount $\Delta
(r,\theta,t)=\delta (r)\cos (\omega _{\rm TOP}t+\theta )$ from
resonance. The time dependence in $\Delta $ is introduced to account
for the rotation of the TOP field, and $\delta (r) $ accounts for the
effects of $\omega _{\mathrm{q}}^{^\prime }$ introduced in
Eq. (\ref{equation:omegaqdash}). Using these parameters we can write
the following equation for the population amplitudes of the $m_{F}$
states $a_{2\ldots -2}$

\[
\fl \frac{\partial }{\partial t}\left(
\begin{array}{l}
a_{2} \\ a_{1} \\ a_{0} \\ a_{-1} \\ a_{-2}
\end{array}
\right) =-\frac{\rmi}{2}\left[
\begin{array}{lllll}
0 & \Omega (t) & 0 & 0 & 0 \\ \Omega (t)^{*} & 2\Delta (r,\theta,t) &
\sqrt{\frac{3}{2}}\Omega (t) & 0 & 0 \\ 0 & \sqrt{\frac{3}{2}}\Omega
(t)^{*} & 4\Delta (r,\theta,t) & \sqrt{ \frac{3}{2}}\Omega (t) & 0 \\
0 & 0 & \sqrt{\frac{3}{2}}\Omega (t)^{*} & 6\Delta (r,\theta,t) &
\Omega (t) \\ 0 & 0 & 0 & \Omega (t)^{*} & 8\Delta (r,\theta,t)
\end{array}
\right] \left(
\begin{array}{l}
a_{2} \\ a_{1} \\ a_{0} \\ a_{-1} \\ a_{-2}
\end{array}
\right).
\]

This model ignores trap loss, this is because we estimate that the
average time taken for an untrapped atom to be ejected from the trap
is approximately 3~ms, so the loss rate $\gamma $ ($\approx 0.3$~kHz)
is small compared with typical values for $\Omega $ and $\Delta $ (in
the range 10 to 100~kHz).

\begin{figure}[tbph]
\begin{center}
\includegraphics[width=\textwidth]{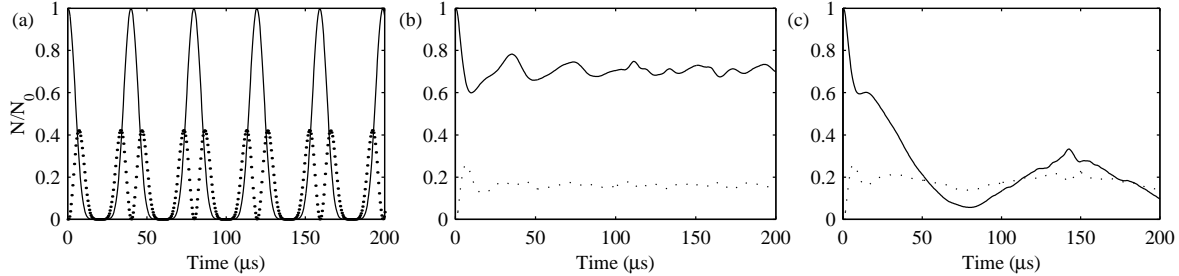} 
\end{center}
\caption{The populations of the $\left| 2,2\right\rangle $ (\full) and
$\left| 2,1\right\rangle $ (\dotted) states as a function of time for
$\Omega = 50$~kHz : (a) homogeneous coupling, (b) inhomogeneous
coupling with very slow bias field rotation and (c) inhomogeneous
coupling with bias field rotation.}
\label{figure:rabicycles}
\end{figure}

Figure \ref{figure:rabicycles} shows the calculated populations of
states $\left| 2,2\right\rangle $ and $\left| 2,1\right\rangle $ as a
function of time for three distinct regimes: (a) strong coupling such
that the coupling rate is uniform across the condensate, (b) very slow
bias field rotation and weak (and therefore inhomogeneous) coupling
across the condensate, and (c) typical experimental conditions of weak
coupling with rapid bias field rotation, which introduces temporal
variation into the coupling. Unless stated otherwise, to match our
experimental conditions we set $\delta (r)=\omega
_{\mathrm{q}}^{^\prime }r$ with $\omega
_{\mathrm{q}}^{^\prime}=14$~kHz~$\mu $m$^{-1}$, and allow a 1~$\mu $s
rise and fall time on $B_{\rm rf}(t)$ which has an amplitude of 72~mG
(equivalent to $\Omega \simeq 50$~kHz).  To calculate the total
population in each state we assume a 3-D Thomas-Fermi distribution of
size 30~$\mu $m~$\times$~30~$\mu $m~$\times$~11~$\mu $m and integrate
over this region.

\begin{figure}
\begin{center}
\includegraphics[width=\textwidth]{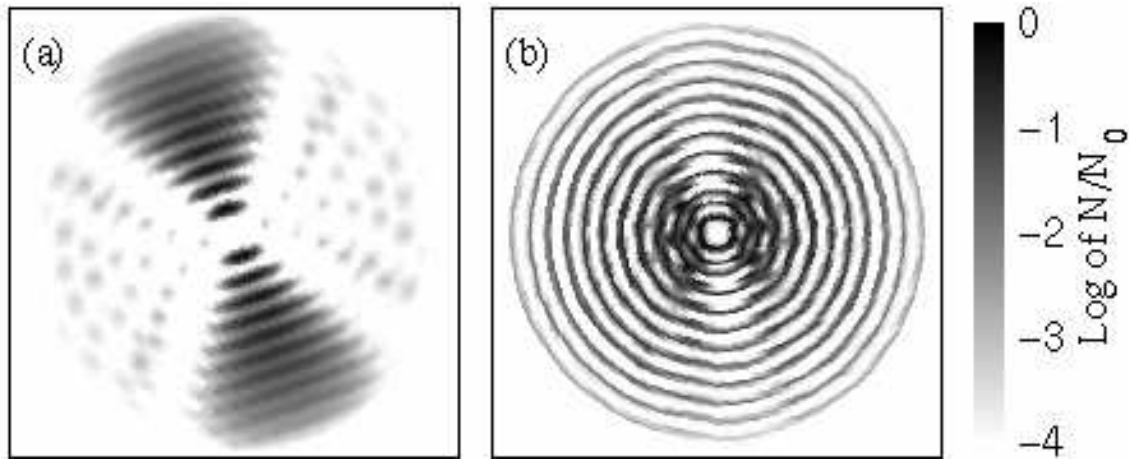}
\end{center}
\caption{Spatial structure of the $\left| 2,2\right\rangle$ state
after RF coupling for (a)~6~$\mu$s and (b)~136~$\mu$s. The condensates
shown are 30~$\mu$m across.}
\label{figure:spatialstructure}
\end{figure}

When the linewidth is significantly larger than the change in
transition frequency across the condensate (as in
Fig. \ref{figure:rabicycles}(a)), spatial effects are reduced and the
population essentially Rabi cycles between the fully stretched states
$\left| 2,2\right\rangle $ and $\left| 2,-2\right\rangle $. For ease
of comparison we keep $\Omega $ constant across all plots in
Fig. \ref{figure:rabicycles} and simulate the broad linewidth
condition in (a) by reducing the quadrupole field gradient by a factor
of 200. For Figs. \ref{figure:rabicycles}(b) and (c) we return to
conditions more relevant to our experiment. For Fig. \ref
{figure:rabicycles}(b), although we have eliminated bias field
rotation, complex spatial behaviour results since the coupling
linewidth is smaller than the Zeeman shift across the
condensate. However the decay of Rabi cycling due to large detuning
changes across the cloud leads to an equilibrium population
distribution that depends only on the pulse linewidth of the
transition relative to the Zeeman shift across the condensate. Figure
\ref {figure:rabicycles}(c) has identical parameters to Fig. \ref
{figure:rabicycles}(b) except that the rotation of the bias field has
been included, so that the time dependence of $\Delta $ is
significant. In this case the Rabi cycling is rapidly overwhelmed by a
population oscillation at the bias field rotation frequency. This
cycling appears only when $r_{0}\omega _{\mathrm{q}}^{^\prime }\gg
\Omega \gg \omega _{\rm TOP}$ where $r_{0}$ is the size of the
condensate, and can be thought of as a frequency modulation of the
coupling field with a beat note at the bias field rotation frequency,
leading to slow cycling of the population. Along with this cyclic
behaviour, our modeling shows that the coupling also introduces
interesting small scale spatial structure in the condensate such as
shown in Fig. \ref{figure:spatialstructure}.

\section{Experimental results}
The focus of this experimental work is to understand the
spatio-temporal dynamics of RF coupling in a TOP trap. To test the
behaviour predicted by the model we first produce Bose condensates in
the $\left| 2,2\right\rangle $ state as described earlier, and then by
varying the duration and power of the RF coupling field we hoped to
obtain results similar to those in Fig. \ref{figure:rabicycles}(b) and
(c).

\begin{figure}[tbph]
\begin{center}
\includegraphics[width=0.8\textwidth]{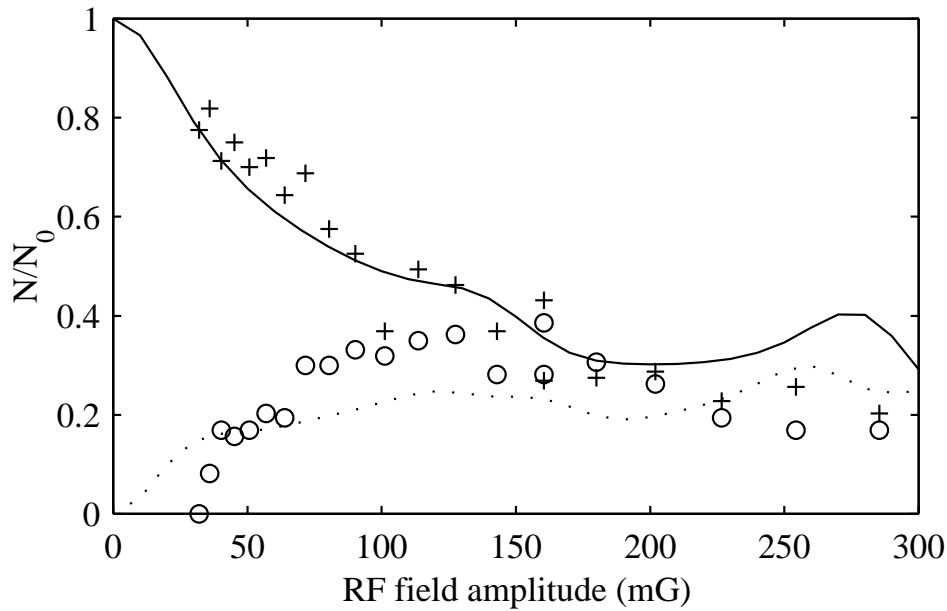}
\end{center}
\caption{Populations of the $\left| 2,2\right\rangle $ (+) and $\left|
2,1\right\rangle $ (o) states as a function of the applied RF magnetic
field. The lines are the results of simulations described in the
text. The error in the data is of order $\pm 10\%$ of N$_{0}$.}
\label{figure:pop_v_power}
\end{figure}

We first considered the case of a relatively short RF pulse, since
this minimises the effect of bias field rotation on the coupling
surface. To do this we investigated the relative populations of the
$\left| 2,2\right\rangle $ and $\left| 2,1\right\rangle $ states as a
function of RF power for a fixed pulse length of 20~$\mu $s (compared
with a trap rotation period of 143~$\mu $s). The results of this,
along with a simulation, are shown in
Fig. \ref{figure:pop_v_power}. It should be noted that there are no
free parameters in any of the simulations of relative population.

The agreement between the experimental and theoretical results is
surprisingly good given the simplistic nature of the model. At a field
amplitude of 150~mG we obtain an approximately equal proportion of
atoms in the two trapped states. For comparison we note that this
field is significantly larger than that typically required for
evaporative cooling.  Within the limits of experimental uncertainty,
the decrease of the $\left| 2,2\right\rangle $ state population with
increased RF field behaves as we might expect, whilst the $\left|
2,1\right\rangle $ state population appears to grow more rapidly than
predicted. The reason for this minor discrepancy may be due to the
quadratic Zeeman effect which would inhibit cycling into the untrapped
states.

\begin{figure}[tbph]
\begin{center}
\includegraphics[width=0.8\textwidth]{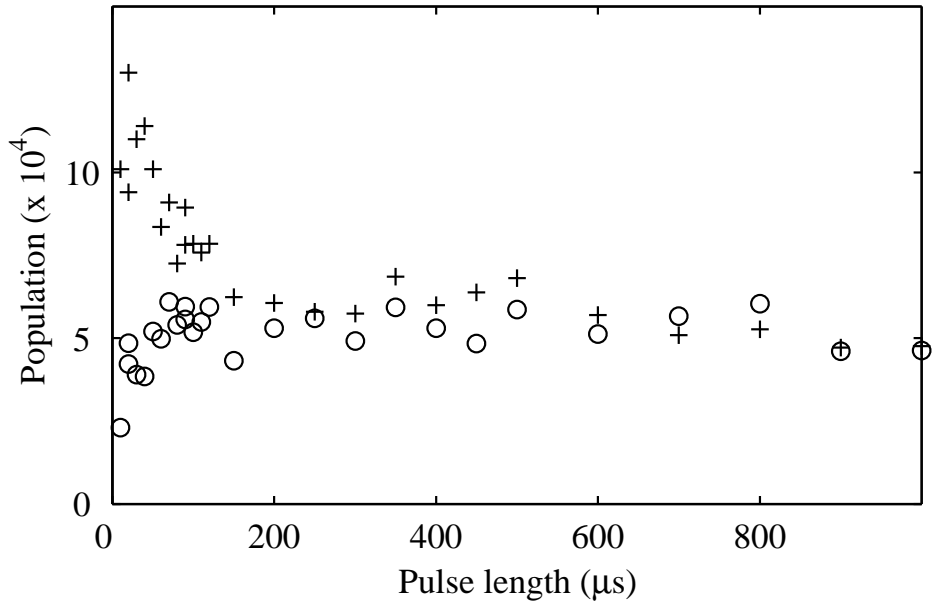}
\end{center}
\caption{Populations of the $\left| 2,2\right\rangle $ (+) and $\left|
2,1\right\rangle $ (o) states as a function of pulse length. The error
in the data is of order $\pm 10\%$.}
\label{figure:pop_v_time}
\end{figure}

Secondly, we considered the case of a weak pulse, since this reduces
the effect of power broadening on the linewidth. To do this we
investigated the relative populations of the two trapped states as a
function of RF pulse length for an RF field amplitude of 72~mG. We
varied the pulse length from 10~$\mu $s to 1~ms to cover a range of
times relative to the bias field rotation period. Results are shown in
Fig. \ref{figure:pop_v_time}. The parameters match those used to
generate Fig. \ref{figure:rabicycles}(c).  Rabi cycling is suppressed
as expected, and the relative populations rapidly become constant and
equal. However, the data does not exhibit the cyclic behaviour of the
simulation and we also observe a linear increase in internal energy of
the condensate as a function of RF pulse length. We believe that both
these observations stem from the spatial structure introduced by the
coupling field. There are large phase gradients associated with the
predicted spatial structure and these correspond to increased internal
energy. The simulation does not include diffusion effects, and atomic
movement on the scale of the spatial structure will tend to average
out the slow population cycling at the bias field rotation frequency.

\section{Conclusions}
To conclude, we have characterised the RF coupling process for Bose
condensates in a TOP trap and shown that it is possible to engineer
predetermined multi-state Bose condensate mixtures. A simple model
predicts the essential features of the condensate behaviour. Our
simulations indicate that the coupling process introduces complex
spatial features. Preliminary results using a full GPE calculation
indicate that simple RF coupling in a TOP trap can lead to vortex
formation \cite{BallaghPC}. This will be the subject of future work at
Otago.

\ack We would like to thank Associate Professor R. J. Ballagh and
Professor W. J Sandle for their helpful advice. We are grateful for
the financial support of the Royal Society of New Zealand Marsden Fund
(contract UOO608), the Foundation for Research, Science and Technology
Postdoctoral Fellowship programme (contract UOO524) and the University
of Otago Research Committee.

\end{document}